# Quasi-phase-matching of only even-order high harmonics

Tzvi Diskin and Oren Cohen

*Solid state institute and physics department, Technion, Haifa, Israel 32000*

We propose a method for generating only even-order high harmonics. We formulate a condition for a shift in the relative-phase between bi-chromatic drivers that leads to a field sign-flip of only even-order harmonics. Induction of this sign-flip periodically during propagation gives rise to quasi phase matching of only even harmonics. We demonstrate this technique numerically and also show that it leads to attosecond pulse trains with constant carrier envelop phase with high repetition rate. This work opens the door for quasi-phase matching of high harmonics with a designed selective enhancement function.

High harmonic generation (HHG) of visible and infrared laser pulses in gases is a useful process for production of coherent extreme ultraviolet and soft x-ray radiation from a tabletop system [1,2]. Applications of HHG include production of attosecond pulse trains (APT) [3], isolated attosecond pulses [4], ultrafast holography [5], coherent diffractive imaging [6], and more. The most common selection rule in HHG from gases media is the absent of even harmonics [2]. This feature reflects the inversion symmetry of the gas and the half-wave symmetry of the driver. Indeed, if the molecules in the gas are oriented [7] or if the half-wave symmetry of the driver is broken [8-11] then the HHG spectrum consists of both odd and even harmonics. It is natural to ask if an HHG experiment can produce the even-order harmonics without the odd-order ones. This conceptual problem was first confronted in Ref. 12 which proposed a coupling geometry of anisotropic quantum dots that can generate terahertz high harmonics where the odd and even harmonics are polarized perpendicularly. This approach cannot be implemented for HHG from gases. In addition to the fundamental science interest in production of pure even harmonics spectra, it may also be useful for applications because a set of phase-locked even harmonics correspond to APT at high repetition rate and stable carrier envelop phase (CEP) (all previously proposed and demonstrated techniques for CEP stabilization of APT reduce the APT repetition rate [13-15].)

Here we suggest a scheme for generation of only even-order harmonics which is based on quasi phase matching (QPM). As in other optical nonlinear processes, HHG can be divided to a regime in which it is phase matched and a regime in which it suffers from phase mismatch [16]. Several QPM techniques have been developed in order to enhance the HHG conversion efficiency in the phase-mismatch regime [17-26]. QPM techniques amplify a spectral region, yet selective control within that region was not obtained. All-optical QPM techniques employ additional weak field in order to coherently control the re-colliding and radiating electronic wave-functions [18,21-24]. The weak driver slightly modifies the electronic trajectories (e.g. by changing the recombination time with attosecond resolution), giving rise to a controlled phase-shift in the phase of the emitted harmonics. Properly designed modulations of the phase-shifts with periodicity that corresponds to two coherence length of the HHG process can lead to efficient QPM.

Here, we propose all-optical QPM of only even-order high harmonics, within a spectral region that include more than 10 harmonics. Both odd and even order high harmonics of a fundamental driver are generated in isotropic and homogeneous media when a secondary driver breaks the half-wave symmetry of the joint pump field. We formulate a condition for a shift of the relative-phase between bi-chromatic drivers that leads to a sign-flip in the fields of only even-order harmonics. Induction of this sign-flip periodically during propagation gives rise to QPM of only even-order harmonics. We demonstrate numerically QPM of only even-order plateau or cutoff harmonics using ti:sapphire pump and its second harmonic weak field that propagate in a dispersive medium. We also numerically demonstrate QPM of even harmonics using weak static field which can be approximated using $CO_2$ or terahertz pulses. Finally, we show that the generated APT exhibits constant CEP and that it consists of two pulses per pump cycle.

In harmonics generation from isotropic and time-independent nonlinear medium, the emitted harmonics field exhibits the same dynamical symmetry as the driving field. For example, a quasi-monochromatic driver field, $E_D$, at angular frequency $\omega_0=2\pi/T$, where T is the optical cycle, is half-wave symmetric: $E_d(t+T/2)=-E_d(t)$, hence the harmonics field, $E_{HHG}$, exhibits the same symmetry: $E_{HHG}(t+T/2)=-E_{HHG}(t)$. The spectrum of this field consists of only odd harmonics of $\omega_0$ because symmetry dictates that even Fourier components of half-wave symmetric functions are zero. The HHG spectrum can include even-order harmonics if a secondary field breaks the fundamental driver half-wave symmetry. This concept was implemented in many experiments where HHG was driven by bi-chromatic drivers that consist of a strong pump and its second harmonic [8]. Also, HHG spectra include both odd and even harmonics of $\omega_0$ when a weak static field (or a very long-wavelength field) is added to the main strong pump [27].

We first present a new symmetry feature for harmonics that are generated by bi-chromatic drivers. We will later employ this feature for QPM of only even-order harmonics. Consider bi-chromatic drivers $E_{BC}=A_0\cos(\omega_0 t+\varphi_0)+A_1\cos(\omega_1 t+\varphi_0+\Delta\varphi)$ where $\omega_0=2\pi/T_0$ and $\omega_1=2\pi/T_1$ are angular optical frequencies, $T_0$ and $T_1$ are optical cycles, $A_0$ and $A_1$ are real amplitudes, $\varphi_0$ is a global phase, and $\Delta\varphi$ is the relative phase between the two components. We compare between the harmonic fields driven by the bi-chromatic fields with the following relative phases: $\Delta\varphi_a=0$ and $\Delta\varphi_b=\pi(1-\omega_1/\omega_0)$. We assign the generated harmonic fields by $E_{HHG}^{a}(t)$ and $E_{HHG}^{b}(t)$, respectively. It is straight forward to verify that $E_{BC}(t,\Delta\varphi_a)=-E_{BC}(t+T_0/2,\Delta\varphi_b)$. The harmonics fields should also conform to this symmetry, hence

$$E_{HHG}^{a}(t) = -E_{HHG}^{b}(t+T_0/2) \qquad (1)$$

Inserting the Fourier decomposition of the emitted harmonic field, $E_{HHG}^{a,b}(t) = \int E_q^{a,b} e^{-iq\omega_0 t} dq$, into Eq. (1) leads to

$$E_q^{a} e^{-i\pi q} = -E_q^{b} \qquad (2)$$

Equation (2) shows that the odd-order harmonics of the bi-chromatic drivers are invariant to a $\pi(1-\omega_1/\omega_0)$ phase-shift of the relative phase, while at the same time, the sign of the even-order harmonics is flipped. This feature is the source for our proposal for QPM of only even-order harmonics. Here, we explore numerically two specific configurations for the bi-chromatic drivers where in both cases the strong pump corresponds to a ti:sapphire laser pulse with central frequency $\omega_0=2.3\times10^{15}$Hz. In the first case, the secondary driver is at much smaller frequency than the pump $\omega_1\ll\omega_0$ (e.g. terahertz or $CO_2$ laser) such that within the pulse-duration of the strong pulse, the field is approximately constant. Numerically, we use a static field for this case. In the second case, the second driver is the second harmonic of the strong pump. We get $\Delta\varphi_b=\pi$ for both cases which corresponds to a change in the sign of the static or second harmonic fields.

Having found a symmetric feature that distinguishes between odd and even harmonics, we now explore it numerically for two specific examples that we will later employ for QPM. In our numerical calculations, we apply the single effective electron approximation and solve the emitted harmonics from singly-ionized xenon (second ionization potential is $I_p$=21 eV) using one-dimensional time-dependent Schrodinger (1D TDSE) solver. In the first case, the bi-chromatic drivers are $E_{BC} = \sqrt{I_0} A_0(t) \cos(\omega_0 t) + E_{DC}$ where $I_0 = 6 \times 10^{14} W/cm^2$, $A_0(t) = \exp\left[-(2t/\tau)^{10}\right]$, $\tau = 50 fs$ is the pulse duration, and $E_{DC}$ is the amplitude of the static field. The cutoff frequency of the HHG spectrum corresponds to the 87$^{th}$ harmonic of the strong pump. Figures 1(a-c) display the emitted phase of several harmonics order as a function of the static field. As expected from our symmetry feature, the phases of even-order harmonics, both at the cutoff and plateau spectral regions, are flipped by π when the static field changes sign. The phases of odd harmonics, on the other hand, do not exhibit such a flip (Fig. 1(c)). Figures 1(d-f) show the intensity of the harmonics as a function of the static field. As shown, the strength of the even harmonics at $E_{DC} \sim 2 \times 10^6$ V/cm is comparable to the strength of odd harmonics without static field. In the second case, we used a bi-chromatic driver of $E_{BC} = A_0(t)\left[\sqrt{I_0}\cos(\omega_0 t) + \sqrt{I_1}\cos(2\omega_0 t + \Delta\varphi)\right]$ where we use three different values for the peak intensity of the secondary field: $I_1 = 0.6 \times 10^{11} W/cm^2, 2.4 \times 10^{11} W/cm^2$ and $9.6 \times 10^{11} W/cm^2$. Figures 2(a-c) display the emitted phase of several harmonics order as a function of the relative phase. As expected from Eq. (2), the field of even harmonics flip their sign (acquire a π phase shift) as a result of a π-shift in the relative phase. At the same time, the phases of odd harmonics are quite constant [Fig. 2(c)]. Notably, within the range shown in Figs. 2(a-c), the variations of the harmonic phases are largely insensitive to the intensity of the second harmonic. Figures 2(d-f) show the intensity of the harmonics as a function of the relative phase. As shown, the strength of even harmonics is in the same order of magnitude as the strength of even harmonics.

Next, we employ the symmetry feature of HHG driven by bi-chromatic drivers for demonstrating numerically QPM of only even-order harmonics in a gas of singly-ionized xenon ions and their free electrons [28]. The strong driver component is a ti:sapphire laser pulse (central wavelength is 0.8 μm) that is initially in the form of $E_0(z=0,t) = A_0(t)\cos(\omega_0 t)$ where $A_0(t) = \sqrt{I_0}\exp\left[-(2t/\tau_0)^4\right]$, $\tau_0 = 36 fs$, $I_0 = 6\times 10^{14} W/cm^2$ and it propagates in z direction (we used super-Gaussian pulse only because it is then more visible in Fig 4 that the produced APT exhibits stable CEP. We verified that Gaussian and hyperbolic secant drivers can also be used for QPM of only even order harmonics and generation of APT with stable CEP.) In the first scheme, the secondary driver is a static field that flips its sign every propagation distance $d_{DC}$: $E_{DC}(z) = g(z)\cdot 2.6\times 10^6 V/cm$ where g(z)=±1 and g(z+$d_{DC}$)=-g(z). This scheme can be implemented experimentally using the setups proposed in Ref. 23. We simulated the propagation of the driver and harmonic fields using the one dimensional version of the model presented in Ref. 29. The nonlinear evolution of the strong driver in the moving frame of light velocity in vacuum, c, is given by:

$$\frac{\partial E_0}{\partial z} = -\frac{1}{2c}\int_{-\infty}^{\tau}\omega_p^2 E_0 d\tau - \frac{2\pi I_P}{(E_0+E_{DC})c}\frac{\partial n_e[E_0+E_{DC}]}{\partial \tau} \qquad (3)$$

where τ=t-z/c, $\omega_p = \sqrt{4\pi e^2 n_e/m}$ is the plasma frequency, where e and m are the electron charge and mass, respectively. The density of free electrons, $n_e$, takes into account the pre-formed plasma and the ionization that is calculated by using the ADK model [30]. The high-order polarization, $P_{HHG}$, is calculated through numerical calculation of the 1D TDSE under the influence of the total field $E_0+E_{DC}$. The generation and evolution of the HHG field, $E_{HHG}$, is described by:

$$\frac{\partial E_{HHG}}{\partial z} = -\frac{2\pi}{c}\frac{\partial P_{HHG}}{\partial \tau} \qquad (4)$$

Figure 3a shows the HHG spectrum after propagation distance of 0.5 mm with gas pressure of 25 torr when $d_{DC}$=18 μm which corresponds to the coherence length of the 88th cutoff harmonic (coherence lengths were calculated for the processes driven by the fundamental field only). For comparison, the generated spectrum with constant static field is also presented. A clear QPM enhancement is obtained around the 88$^{th}$ harmonic. Figure 3b shows the coherent buildup of the 88$^{th}$ and 87$^{th}$ harmonic fields, showing clearly that the odd harmonic experience a QPM enhancement while the even harmonic suffers from phase-mismatch. Notably, the QPM efficiency of the 88$^{th}$ harmonic is 0.27, which is relatively high for QPM in HHG [31]. Figure 3c shows the HHG spectrum that is generated when $d_{DC}$=28 μm which corresponds to the coherence length of the 70$^{th}$ plateau harmonic. Clear QPM enhancement is obtained around the 72$^{th}$ harmonic. Figure 3d shows the coherent buildup of the 70$^{th}$ and 71$^{th}$ harmonic fields, showing again that the even harmonic experience a QPM enhancement (with 0.23 QPM efficiency) while the odd harmonic suffers from phase-mismatch. The generated even-harmonics correspond to high repetition-rate APT with stable CEP. This feature is demonstrated in Fig. 4. Fig 4a shows the normalized APT, $E_{QPM}(t)$ that corresponds to the red spectrum in Fig. 3a in the spectral region 83±5 harmonics. Figure 4b shows the average of $E_{QPM}$ and its $T_0/2$ time-delayed, showing that this APT has a stable CEP. Notably, the temporal distance between consecutive pulses is $T_0/2$. That is, in contrast to previous methods [13-15], APT with stable CEP is obtained without reduction of the repetition rate. For comparison, Fig. 4c shows $E_{SA}$ (SA stands for single atom) which corresponds to the APT generated by the same strong pump beam, but without propagation and without the static field. The average of $E_{SA}$ and its $T_0/2$ time-delayed show that consecutive pulses have opposite phases (Fig 4d).

Next, we demonstrate numerically QPM of only even-order harmonics when the secondary driver is the second harmonic of the strong pump. We assume that the second harmonic field experiences an effective refractive index that is Δn smaller than the refractive index of the strong pump. This scenario can be implemented experimentally by using highly dispersive nonlinear medium [28], or by utilizing spatial dispersion in hollow planar waveguide [32], as proposed in Ref. 22. As a result of the dispersion, the relative phase between the drivers evolves during propagation, and after some propagation distance, $L_π$, it acquires a π shift. QPM is obtained if this distance corresponds to the coherence length of the process, $L_\pi=L_C$. The incident beam in our simulation is $E_{BC}=E_0+E_1$ where $E_0$ is the same as in the previous section and $E_1 = \sqrt{I_1}\exp\left[-(2t/\tau_1)^4\right]\cos(2\omega_0 t)$ where $\tau_1 = 72\,fs$ and $I_1 = 2.4\times 10^{11} W/cm^2$. We simulated the propagation of the beam using the following equation:

$$\frac{\partial E_{BC}}{\partial z} = -\frac{1}{2c}\int_{-\infty}^{\tau}\omega_p^2 E_{BC}d\tau - \frac{2\pi I_P}{Ec}\frac{\partial n_e E_{BC}}{\partial \tau} - \frac{1}{c}\frac{\partial}{\partial \tau}\int_{-\infty}^{\infty}\Delta\tilde{n}(t')E_{BC}(\tau-t')dt' \qquad (5)$$

where $\Delta\tilde{n}(t')$ is the inverse Fourier transform of the dispersion $\Delta n(\omega)$, which is zero at the spectral region near $\omega_0$ and $-\Delta n$ in the region around $2\omega_0$. The third term in Eq. (5) gives rise to the assumed dispersion, only. Figure 5a shows the HHG spectrum when $\Delta n=8.7\times 10^{-3}$ ($L_\pi=46$ μm) and after propagation distance 1 mm. For compression, the generated spectrum when $\Delta n=0$ is also presented. A clear QPM enhancement is obtained around the 86$^{th}$ harmonic. Figure 5b shows the coherent buildup of the 86$^{th}$ and 85$^{th}$ harmonic fields, showing clearly that the even harmonic experience a QPM enhancement (QPM efficiency is 0.27) while the odd harmonic suffers from phase-mismatch. Figure 5c shows the HHG spectra when $\Delta n=7\times 10^{-3}$ ($L_\pi=57$μm) and, for compression also the $\Delta n=0$ case. A clear QPM enhancement is obtained around the 70$^{th}$ harmonic. Figure 5d shows the coherent buildup of the 70$^{th}$ and 71$^{th}$ harmonic fields, showing that the even harmonic experience a QPM enhancement (QPM efficiency is 0.14) while the odd harmonic suffers from phase-mismatch.

In conclusions, we propose and demonstrated numerically a technique for generating only even-order harmonics, within a spectral region that contain ~10 harmonics, using quasi-phase matching. This technique shows that symmetry arguments can be employed for selective control over the spectral features of HHG.

**FIGURE CAPTIONS**

**Figure 1:** Influence of secondary static field on the phases and intensities of even and odd harmonics. Phases (a-c) and intensities (d-f) of the $86^{th}$, $70^{th}$ and $85^{th}$ harmonics as a function of the static field. The phases of even and odd harmonics are odd and even functions of the static field, respectively.

**Figure 2:** Influence of the relative phase between the fundamental and its second harmonic drivers, $\Delta\varphi$, on the phases and intensities of even and odd harmonics. Phases (a-c) and intensities (d-f) of the $84^{th}$, $60^{th}$ and $85^{th}$ harmonics as a function of the relative phase for different intensities of the second harmonic field: $0.6\times10^{11} W/cm^2$ (blue), $2.4\times10^{11} W/cm^2$ (red) and $9.6\times10^{11} W/cm^2$ (black). A $\pi$ shift in the relative phase results with a $\pi$-shift in phases of even harmonics (sign flip of their fields) and no shift in phases of odd harmonics.

**Figure 3.** Numerical demonstration of QPM of only even harmonics using a periodic static field. (a) Harmonic spectra driven by a fundamental driver and a static field with sign (direction of polarization) that is flipped periodically during propagation with periodicity that corresponds to the coherence length of the $88^{th}$ cutoff harmonic (red) and with static field with a constant sign (black). The inset shows the spectra near the $88^{th}$ harmonic, showing that only even order harmonics are enhanced. (b) Buildup of the $88^{th}$ harmonic (red), the $87^{th}$ harmonic with periodic static field (dashed blue) and the $88^{th}$ harmonic with constant static field (dotted black). The buildup of the $88^{th}$ harmonic with periodic static field exhibits a typical QPM structure. (c) and (d) show the same

data as (a) and (b), respectively, with the only difference that the periodicity of the static field corresponds to the coherence length of the 70$^{th}$ plateau harmonic.

**Figure 4:** Generation of APT with stable CEP without reducing the repetition rate. (a) Normalized APT, $E_{QPM}(t)$, corresponds to the red spectrum in Fig. 3a in the spectral region 83±5 harmonics. (b) Averaged sum of $E_{QPM}$ and its $T_0/2$ time-shifted field. The fact that the average sum is very similar to $E_{QPM}$ shows that consecutive pulses are very similar, i.e they all have the same CEP. (c) Normalized APT, $E_{SA}(t)$, corresponds to the spectral region 83±5 harmonics of a spectrum generated by only the fundamental driver and without propagation. (d) Averaged sum of $E_{SA}$ and its $T_0/2$ time-shifted field. The fact that the central pulses of the average are almost zero corresponds to the known fact that consecutive pulses in ordinarily generated APT have opposite phases.

**Figure 5:** Numerical demonstration of QPM of only even harmonics using a bi-chromatic drivers that consists of a fundamental pump field and its second harmonic. (a) Harmonic spectra when the two drivers experience dispersion $\Delta n=8.7\times10^{-3}$ (red) and $\Delta n=0$. The inset shows the spectrum near the 86$^{th}$ harmonic, showing that it consists of only even order harmonic. (b) Coherent buildups of the 86$^{th}$ (red) and the 85$^{th}$ (dashed blue) with $\Delta n=8.7\times10^{-3}$ and the 86$^{th}$ harmonic with $\Delta n=0$ (dotted black). The 86$^{th}$ harmonic $\Delta n=8.7\times10^{-3}$ rises much more rapidly than the other buildup curves. (c) Harmonic spectra when the two drivers experience dispersion $\Delta n=7\times10^{-3}$ (red) and $\Delta n=0$ (black). The inset shows the spectrum near the 70$^{th}$ harmonic, showing that it consists of only even order harmonic. (d) Coherent buildups of the 70$^{th}$ (red) and the 71$^{th}$ (dashed blue) with $\Delta n=7\times10^{-3}$ and the 70$^{th}$ harmonic with $\Delta n=0$ (dotted black). The 70$^{th}$ harmonic $\Delta n=7\times10^{-3}$ rises much more rapidly than the other buildup curves.

**Figure 1**

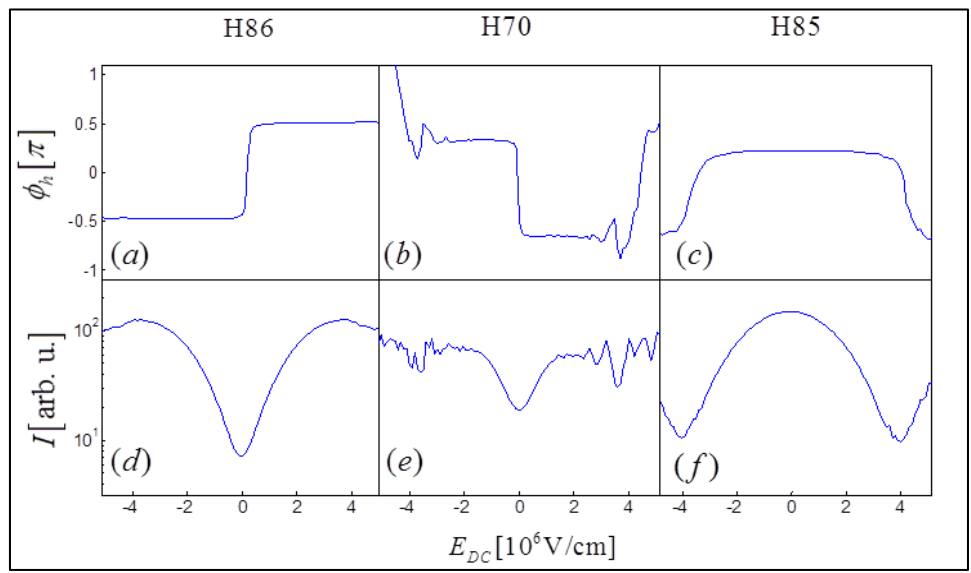

**Figure 2**

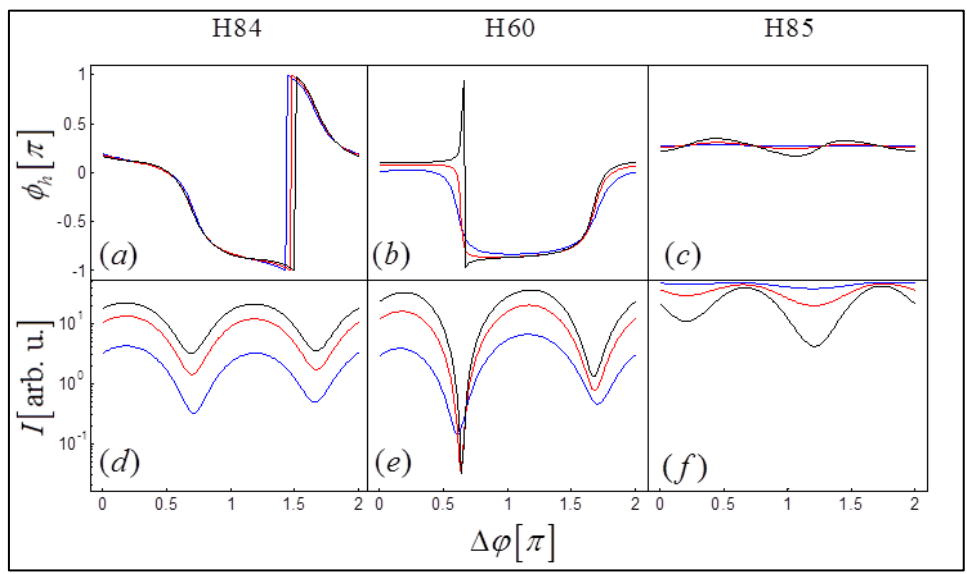

**Figure 3**

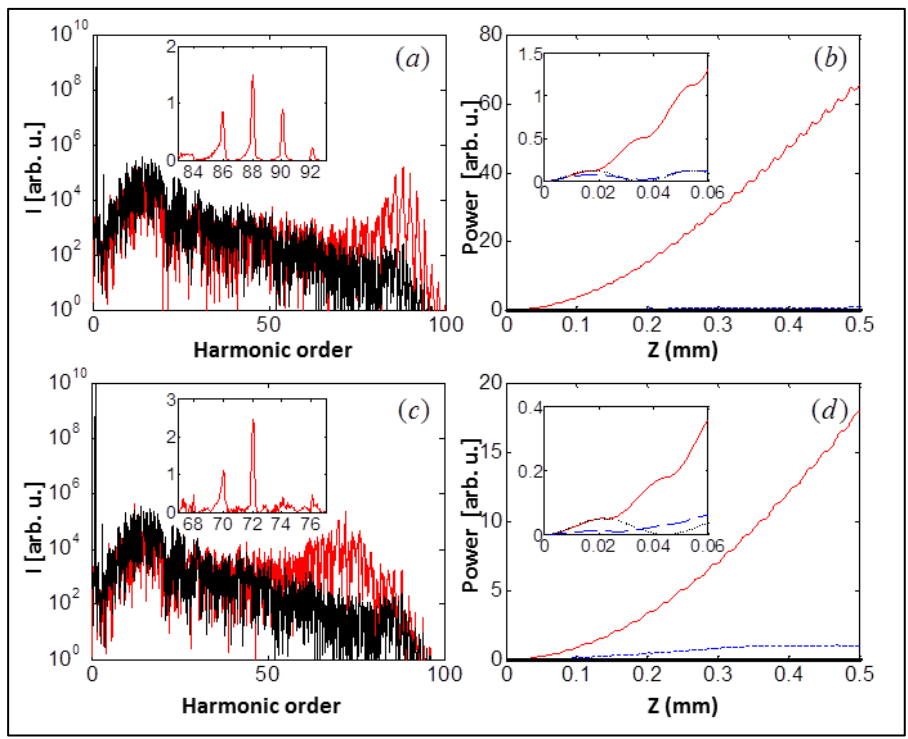

**Figure 4**

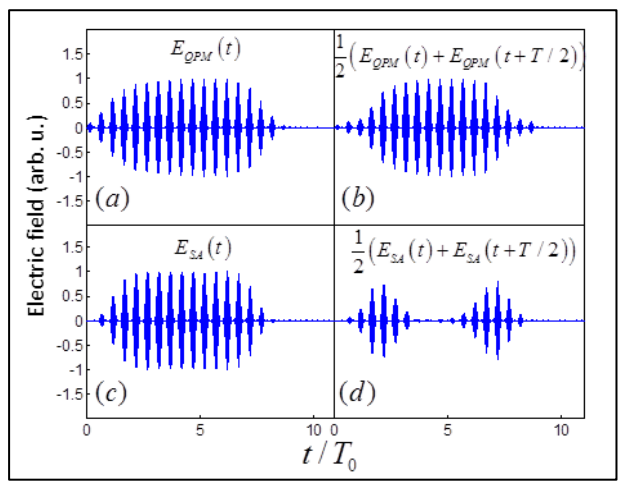

**Figure 5**

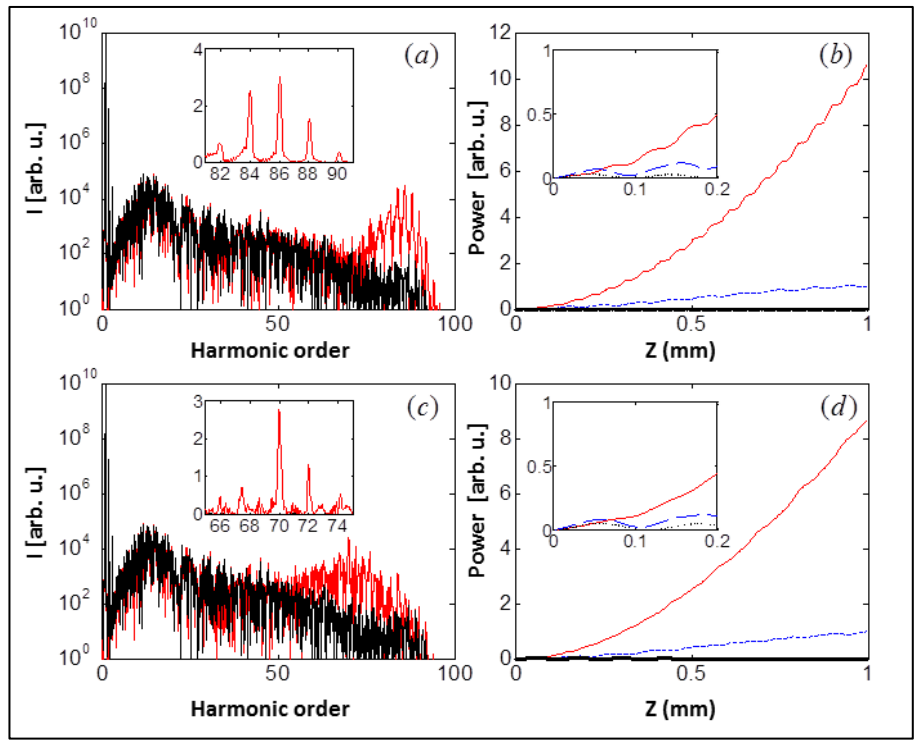